\documentclass[submission,copyright,creativecommons]{eptcs}
\providecommand{\event}{ACL2 Workshop 2022} 
\usepackage{underscore}           
\usepackage{algorithm}
\usepackage{algorithmic}
\usepackage{longtable}
\usepackage{varwidth}
\usepackage{verbatim}

\newenvironment{centerverbatim}{%
  \par
  \centering
  \varwidth{\linewidth}%
  \verbatim
}{%
  \endverbatim
  \endvarwidth
  \par
}

\title{A Mechanized Proof of Bounded Convergence Time \\ for the Distributed Perimeter Surveillance System (DPSS) Algorithm A}
\author{David Greve
  \institute{Collins Aerospace}
  \and
  Jennifer Davis
  \institute{Collins Aerospace}
  \and
  Laura Humphrey
  \institute{Air Force Research Laboratory}
}

\def\titlerunning{DPSS-A Convergence}
\def\authorrunning{Greve, Davis \& Humphrey}
\begin{document}
\maketitle

\begin{abstract}
The decentralized perimeter surveillance system (DPSS) seeks to
provide a decentralized protocol for evenly distributing surveillance
of a perimeter over time across an ensemble of unmanned aerial
vehicles (UAVs) whose members may communicate only when in close
proximity to each other.  The protocol must also converge to an even
distribution of the perimeter in bounded time.  Two versions of the
DPSS protocol presented in \cite{DPSS} seem to converge in bounded time but
only informal proofs and arguments are given.  A later application of
model checking to these protocols found an error in one of the key
lemmas, invalidating the informal proof for one and casting doubt on
the other \cite{Intuit}.  Therefore, a new hand proof of the convergence time
for the simpler version of the DPSS protocol or algorithm, Algorithm A
or DPSS-A, was developed by Jeremy Avigad and Floris van Doorn \cite{AvD}.
This paper describes a mechanization of that hand proof in the logic
of ACL2 and discusses three specific ACL2 utilities that proved useful
for expressing and reasoning about the DPSS model.
\end{abstract}

\section{Introduction}

The decentralized perimeter surveillance system (DPSS) seeks to
provide a decentralized protocol for evenly distributing surveillance
of a perimeter over time across an ensemble of unmanned aerial
vehicles (UAVs) whose members may communicate only when in close
proximity with each other.  The protocol must also converge to an even
distribution of the perimeter in bounded time.  The original DPSS
paper presents two protocols or algorithms for solving this problem:
Algorithm A and Algorithm B.  Algorithm B is the more general
algorithm, supporting perimeters whose lengths may change and
ensembles of UAVs that may add or lose members over time.  Algorithm A
is much simpler, assuming that the perimeter and the number of UAVs is
fixed.  This work focuses almost exclusively on DPSS Algorithm-A,
which we refer to as DPSS-A.

In general, designing protocols for multi-agent interaction that
achieve the desired behavior is a challenging and error-prone
process. A common practice, as in the original work on DPSS, is to
manually develop non-mechanized proofs of protocol correctness that
rely on human intuition and require significant effort to
develop. However, even given a high level of effort, such proofs can
have mistakes that may go unnoticed after peer review, modeling and
simulation, and testing, motivating the need for proof mechanization.

Indeed, previous efforts to formally verify bounds on DPSS convergence
have already realized the benefits of proof mechanization.  In \cite{Intuit},
Davis and Humphrey mechanized proofs of convergence for concrete
instances of the DPSS algorithm using the AGREE model checker.  The
original conjecture was published nearly ten years prior, had received
close to 200 citations, and provided a compelling ``proof'' of correctness
backed by extensive simulation results.  Mechanized analysis by the
model checker, however, found a counterexample to a key lemma that
demonstrated that the earlier ``proof'' of correctness was, in fact,
incorrect.  The counterexample revealed a key intuition that was
missing and identified a corner case in an assumption about the
initial conditions that invalidated the proof of Algorithm B, which
also cast doubt on the correctness of the proof of Algorithm A.
Overall, the model checking results suggest that both algorithms
converge in bounded time.  However, while model checking can analyze
individual instantiations of the system, i.e. ensembles with a
specified number of UAVs, it cannot verify convergence times for an
arbitrary number of UAVs.  For this we need the expressiveness of a
theorem prover.

This report focuses on efforts to mechanize, in the logic of the ACL2
theorem prover, a new hand proof developed by Jeremy Avigad and Floris
van Doorn that bounds the convergence time of DPSS-A for an arbitrary
number of UAVs.  Our mechanized proof provides assurance that there
are no unfounded intuitions in the hand proof and provides confidence
that we have captured and addressed all essential assumptions and
corner cases related to the DPSS algorithm.  Furthermore, for each of
the two essential invariants employed in the hand proof, we provide
concise, useful specifications relative to a concrete formalization of
the DPSS-A algorithm.  Finally, we discuss three ACL2 utilities that
proved useful in formalizing and reasoning about our DPSS model.


\section{DPSS Overview}

The problem underlying DPSS is as follows.  Suppose you have a
perimeter mapped to a line segment and an ensemble of UAVs that
surveil the perimeter as they move along it, each moving at the same
uniform speed.  Suppose that the perimeter endpoints and members of
the UAV ensemble are fixed.  Suppose UAVs communicate with each other
if and only if they are in close proximity, reduced to the extreme
case of being co-located (modulo altitude deconfliction).  Finally,
suppose UAVs are able to detect the perimeter endpoints.  Define the
optimal surveillance pattern as one in which the perimeter is divided
into segments of equal length, with one UAV surveilling each segment,
and with each UAV arriving at the endpoint of its segment at the same
time as its neighbor in a synchronized and periodic manner.  Then the
question is whether there is a decentralized protocol whose
steady-state behavior converges to this optimal surveillance pattern
in bounded time.  The motivation for bounded time convergence is to
make the system robust to changes in the perimeter and UAV ensemble.
That is, if the protocol can be guaranteed to re-converge within a
relatively short time bound for a fixed perimeter and UAV ensemble,
and if changes to the perimeter and UAV ensemble are relatively
infrequent, then the system will often be in the optimal surveillance
pattern.

The original work on DPSS \cite{DPSS} proposes a general protocol or algorithm
to solve this problem.  Let us refer to the endpoints of the perimeter
as \emph{left} and \emph{right}.  Let us assign sequentially
increasing IDs to the UAVs starting with the leftmost UAV, assuming
UAVs do not start as co-located or there is some mechanism to break
ties if they do.  Recall the assumptions that UAVs all move at the
same uniform speed, communicate if and only if they are co-located,
and can detect when they have reached the perimeter endpoints.  In the
proposed algorithm, each UAV has set of left and right
\emph{coordination variables} whose values reflect the UAV's
current beliefs about how far it is to the perimeter's left and
right endpoints and how many UAVs are to its left and right.
Initially these may be incorrect.  However, once a UAV reaches the
left perimeter endpoint, it knows its distance to the left endpoint is
0 and the number of UAVs to its left is also 0, and it updates its
left coordination variables to the correct values and changes
direction, updating its distance to the perimeter endpoint as it
moves.  An analogous situation holds for the right.  When two UAVs
become co-located or \emph{meet}, they exchange information, and the
left UAV updates its right coordination variables based on those of
the right UAV and vice versa.  After meeting, the two UAVs compute
where they believe the shared boundary of their segments is,
\emph{escort} each other to that point if they are not already there,
then separate and move along their own segments.  Note that the only
conditions under which a UAV is allowed to change direction are those
previously mentioned: when it reaches a perimeter endpoint, when it
starts escorting another UAV to their shared segment boundary, or
when it separates from its neighbor at a shared segment boundary.

Intuitively, the escort part of the protocol ensures that UAVs remain
sequentially ordered and arrive at their shared segment boundaries at
the same time, and information exchanges about coordination variables
eventually propagate correct values for these variables across all
UAVs.  From an arbitrary initial state, the behavior of this algorithm
may appear chaotic, with some UAVs traversing large portions of the
perimeter before interacting with another UAV. As time progresses,
however, regular patterns begin to emerge. The claim is that after a
finite number of information exchanges and escorts, the UAVs achieve
consensus on correct coordination variable values and also eventually
achieve synchronous and periodic arrival at their shared segment
boundaries, causing each to be confined to its own segment.  In other
words, the UAVs achieve the optimal surveillance pattern in bounded
time.

For analysis and proof purposes, the original work on DPSS
distinguishes between two phases of the protocol or algorithm.
Algorithm A assumes that UAVs start with correct coordination
variables, whereas Algorithm B does not.  In other words, Algorithm B
is the general algorithm, and it reduces to Algorithm A once all UAVs
have correct values for their coordination variable.  This paper
focuses on Algorithm A, which we refer to as DPSS-A, and which is
shown in Figure 1.

\begin{algorithm}
  \caption{DPSS-A}
  \begin{algorithmic}
    \STATE
    \IF{UAV \textbf{i} meets with neighbor \textbf{j}}
    \STATE Travel with neighbor \textbf{j} to shared segment position
    \STATE Set direction to monitor own segment
    \ELSIF{reached perimeter endpoint}
    \STATE Reverse direction
    \ELSE
    \STATE Continue in current direction
    \ENDIF
    \STATE
  \end{algorithmic}
\end{algorithm}

Note that in the original paper, Algorithm A includes a vestigial step
in which UAVs exchange information on coordination variables, even
though all UAVs are assumed to already agree on the correct values.
For the purpose of establishing the convergence of DPSS-A, therefore,
the exchange of coordination variables is irrelevant.  Consequently,
we do not include coordination variables in our UAV model, and the
code for the coordination variable consensus is omitted from our
formalization of the DPSS-A algorithm.

\subsection{Fundamental Data Structures}

The base DPSS-A model provides the mathematical foundation upon which
the Jeremy Avigad and Floris van Doorn (AvD) proof is built.  We
endeavored to make the base model as simple as possible while still
being able to specify the relevant behaviors of DPSS-A. Our model
describes a perimeter of a fixed length and a UAV ensemble consisting
of a fixed but arbitrary number UAVs on that perimeter.  In our model,
constrained functions are used as theory parameters.  This approach
allows us to avoid explicitly threading the global model parameters
through every relevant function.  ACL2 supports functional
instantiation, which would allow the resulting theorems to be
instantiated for specific values of the theory parameters.  The length
of the perimeter is modeled using a nullary function {\tt (P)} constrained
to return a non-zero real (rational) value.  The number of UAVs is
also represented as a nullary function {\tt (N)} that is constrained to be
an integer greater than or equal to 1.  The length of each segment is
defined such that {\tt (S) = (P)/(N)}.  Time and UAV velocity are normalized
so that a UAV may traverse one segment length in one unit of time.
The value of {\tt T} (in the original paper) is the time required for a UAV
to traverse the entire perimeter.  Relative to our model, therefore,
{\tt T = (N)}.

A well-formed UAV ensemble is modeled as a list that contains (N)
UAVs.  Each UAV has a unique identifier which is a natural number in
the range $0$ to {\tt (N)-1} inclusive as well as a position (on the
perimeter) and a direction of travel.  The UAV identifiers increase
sequentially from left to right.  The UAVs are assumed to be ordered
on the perimeter such that if $(i<j)$ then the location of UAV i is
always less than or equal to that of UAV j.  A well-formed UAV
ensemble is recognized by the predicate {\tt (wf-ensemble ens)} which
appears as a hypothesis in many of our proofs.  Finally, we say that
the segment of the perimeter assigned to UAV i starts at location
$i*(S)$ and ends at $(i+1)*(S)$.

\subsection{DPSS-A Event Based Simulator}

We model the behavior of the UAV ensemble using an \emph{event-based
  simulation} of the DPSS algorithm.  In this simulation, time
advances continuously by stepping from one event to the next, though reasoning
about the state of the system between events is supported.  All events
in the simulation correspond to changes in at least one UAV's
direction as described by the DPSS algorithm, which for brevity we
will refer to as \emph{flips}. After an event, consisting of one or
more UAV flips, the location of all the UAVs are updated based on
their direction of motion and the amount of continuous time before the next
event.  The top-level DPSS-A simulator {\tt (step-time dt ens)} allows
this process to be repeated for an arbitrary length of time.

The function {\tt (event-for-uav i ens)} takes the index of a UAV and
an ensemble and returns true if the UAV is currently experiencing an
event.  The function {\tt (flip-on-events ens)} flips the direction of
any UAV in the ensemble currently experiencing an event.  An important
property of {\tt flip-on-events} is that, once all of the UAVs with events
have been flipped, there won't be any remaining events at that
instance in time.  This property relies on an appropriate definition
of {\tt event-for-uav}.  Note that the DPSS-A model assumes UAVs can change
direction instantaneously, as assumed in \cite{DPSS}.  This is convenient in
that there is no need to budget for turning UAVs in the convergence
time bound.  One fall-out of this decision, however, is that a
UAV's direction is discontinuous during events. For example, the
statement ``if we have an event and the UAV is moving left'' can
be both true and false at the same instant in time.  This impacts the
formalization of certain properties.  It also introduces corner cases
that we might otherwise intuitively dismiss, but which must be
addressed in the proof in order to ensure that we are consistent with
the assumptions we are making about the behavior of the system.

The predicate {\tt (impending-impact-event-for-uav i ens)} recognizes the
conditions under which the time to the next impact event for the given
UAV can be computed given only the state of its immediate neighbors
and, for the outermost UAVs, the perimeter endpoints.  If no impact
event is impending, then the UAV must be chasing one of its neighbors
and it won't actually have an event until after that neighbor
experiences an event and changes direction.  An impact event is
defined as either an actual {\tt event-for-uav} or an \emph{escort event}.  On an
actual {\tt event-for-uav}, UAV i will change directions.  For an escort
event, UAV i will not change directions, but one of its neighbors will
(and UAV i will escort it back to their shared segment boundary).
Note that if an impact event is impending for UAV i, then an actual
event is impending for some UAV, either UAV i or one of its neighbors.
The function {\tt (min-time-to-impact-for-uav i ens)} computes the minimum
time before an impact might take place for the i'th UAV based only
on the state of its immediate neighbors and the endpoints.  If an
impact event is impending for UAV i, then {\tt event-for-uav} will be true
in exactly {\tt min-time-to-impact-for-uav} time increments.

The function {\tt (always-smallest-min-time-to-impending-impact ens)} takes
an ensemble and returns the smallest rational time increment that will result in an
actual event for some UAV.  An early DPSS model used the smallest
{\tt min-time-to-impact-for-uav} regardless of whether an actual event was
impending for that UAV.  While this ensured that the time increment was
less than or equal to any time to impact, it did not ensure useful
progress.  This, in turn, complicated subsequent high-level proofs.
By requiring that the simulator advance to the next actual event we
complicated some low-level proofs (for instance, we were required to
prove that, in any ensemble configuration, there is always at least
one UAV with an impending event) but we ultimately simplified the
high-level proofs which, in our estimation, was a good trade-off.

The function {\tt (update-location-all dt ens)} takes a time increment and an
ensemble and returns the ensemble state after advancing time by the
specified time increment.  The function {\tt (next-step dt ens)} takes a
requested time increment and an ensemble.  First it applies {\tt flip-on-events}.
It then compares the requested time increment with the
{\tt always-smallest-min-time-to-impending-impact} and performs an
{\tt update-location-all} with the lesser of the two, returning the
difference between the requested time increment and the applied time increment
along with the new ensemble.

The top-level simulator function is {\tt (step-time dt ens)}.  It takes a
requested time increment and an ensemble.  If the requested time increment is
zero, it returns the ensemble.  If it is non-zero, it calls {\tt next-step}
to compute a new (smaller) time increment and an updated ensemble state and
calls itself recursively.  An important property of {\tt step-time} is that
stepping time by $A$ and then stepping time by $B$ is the same as stepping
time by $A+B$.

\section{AvD Proof}

In \cite{AvD} Jeremy Avigad and Floris van Doorn (AvD) provide a hand proof
of bounded convergence time for DPSS-A.  The formalism used in the AvD
proof differs from our base model primarily in the fact that the AvD
UAV index ranges between 1 and $N$ inclusive (as in the original paper)
and the UAV velocity is normalized to cover the entire perimeter in
one unit of time.  The convergence time for Algorithm A is expressed
in the AvD paper as $2-1/N$ units of time, improving on the previously
reported convergence time bound of $2$, where the improvement can be
traced to a more precise notion of convergence.  Our formalism defines
$T$ as the time required for a UAV to traverse the entire perimeter.
Thus, we express the final bound on convergence time as $2*T - 1$.

The AvD proof employs 7 lemmas that hinge on two fundamental concepts:
\emph{have met} and \emph{synchronized}. The definition of these
concepts, as they appear in the hand proof, are:

\textbf{Have Met} ``We say that two [UAVs] have met by time t if either they
  started together, moving in the same direction, or they have been
  involved in a meet or bounce event.''
  
\textbf{Synchronized} ``A [UAV] is left/right synchronized at time t if
  beyond that point it never goes to the left/right of its left/right
  [segment] endpoint.''

Essentially the claims are that a) all UAVs will \emph{have met} after
time $T$ and b) once all of the UAVs have met, they will all be
synchronized within $T-1$ time increments and c) the synchronization of the
entire ensemble is equivalent to convergence of the algorithm.  The
proof is presented in terms of the more refined concepts of \emph{have
  met (left)} and \emph{left synchronized} and an argument is then
made that the \emph{right} version of the proof follows by symmetry.

One important contribution of our effort was the development of
precise formalizations of the concepts of \emph{have met} and
\emph{synchronized} that are expressed relative to a concrete model of
the DPSS algorithm and formulated in a manner amenable to proof
mechanization.  Expressing these concepts relative to a concrete model
forced us to consider subtle corner cases.  For example, the fact that
a UAV may have two different directions in the same instant of time
required some care when formulating predicates that remained invariant
when UAVs changed direction.  Such corner cases would be easy to
overlook outside the context of a mechanized proof.

Regarding mechanization, it is easier to reason about concepts that
can be expressed \emph{locally} in terms of both time and space.  In
terms of time, we are looking for predicates that take a current state
and recognize interesting behaviors, hopefully enabling us to then
make predictions about the behavior of future states.  In terms of
space, we want predicates that speak of UAVs in isolation or, when
necessary, the relationships between a given UAV and a finite number
of neighbors (i.e., the adjacent UAVs).  The
{\tt impending-impact-event-for-uav predicate}, for example, is expressed
only in terms of the given UAV and its neighbors and the endpoints for
the outermost UAVs.


Early attempts to formalize \emph{have met} and \emph{synchronized},
however, appealed to either a history of DPSS execution steps or an
extrapolation of execution steps into the future or were expressed
\emph{non-locally} in terms of the states of an arbitrary number of
adjacent UAVs.  While not incorrect, such formalisms a) tend to draw
heavily on human intuition about system behavior and b) are often
difficult to work with in a mechanized proof.  After extended discussions
with Avigad and van Doorn we were finally able to articulate refined
notions of these concepts that a) capture the author's intent b) offer
simple, localizable predicates over an arbitrary ensemble of UAVs and
c) are provably invariant relative to the DPSS-A algorithm.  Our
formalization of the \emph{left} version of \emph{have met} is shown
below\footnote{The {\tt def::un} macro is simply a {\tt defun} wrapper provided
  by the the \emph{coi} libraries that supports the specification of function signatures via
  the {\tt xargs} keyword {\tt :fty} and is available via 
  {\tt (include-book "coi/util/defun" :dir :system)}}.

\begin{verbatim}
(def::un have-met-Left-p (i ens)
  (declare (xargs :fty ((uav-id uav-list) bool)))
  (let ((uavi  (ith-uav i ens))
        (right (ith-uav (+ i 1) ens)))
    (implies
      (and
       (< i (+ -1 (N)))
       (< (UAV->direction uavi) 0))
      (and
       (implies
        (< (UAV-right-boundary uavi) (UAV->location uavi))
        (and
         (< (UAV->direction right) 0)
         (equal (UAV->location uavi) (UAV->location right))))
       (implies
        (and (<= (UAV->location uavi) (UAV-right-boundary uavi))
             (<  (UAV-left-boundary uavi) (UAV->location uavi)))
        (and
         (implies
          (< (UAV->location uavi) (UAV-right-boundary uavi))
          (< 0 (UAV->direction right)))
         (equal (average (UAV->location uavi) (UAV->location right))
                (UAV-right-boundary uavi))))
       ))))
\end{verbatim}

In English this says that the rightmost UAV \emph{has met (left)}, as
has any UAV moving to the right.  For any other UAV that is moving to
the left, if it is right of its rightmost segment boundary, the UAV to its
right is escorting it back to their shared boundary.  If the UAV is in
its segment but not on the left boundary, then the UAV is the same
distance from its right segment boundary as the UAV to its right and, if it is
left of its right boundary, then the UAV is moving to the right.

Our formalization of the \emph{left} version of \emph{synchronized}
is shown below.

\begin{verbatim}
(def::un left-synchronized-p (j ens)
  (declare (xargs :fty ((uav-id uav-list) bool)))
  (implies
   (< 0 j)
   (and
    (<= (UAV-left-boundary (ith-uav j ens))
        (average (UAV->location (ith-uav (+ -1 j) ens))
                 (UAV->location (ith-uav j ens))))
    (implies
     (and
      (< (UAV->direction (ith-uav j ens)) 0)
      (not (equal (UAV->location (ith-uav j ens))
                  (UAV->location (ith-uav (+ -1 j) ens)))))
     (< 0 (UAV->direction (ith-uav (+ -1 j) ens)))))))
\end{verbatim}

In English this says that the leftmost UAV is \emph{left
  synchronized} and, for any other UAV, the average of its location with
its left neighbor's location is not less than its left segment
boundary.  Additionally, if it is moving left and it is not co-incident
with its left neighbor, then its left neighbor is moving right.

\subsection{Proof Overview}

After defining \emph{have met (left)} and \emph{left synchronized} we
establish that these predicates are invariant over {\tt step-time}.  While
the invariance of \emph{have met (left)} is straightforward, the
invariance of \emph{left synchronized} depends on \emph{have met
  (left)}.  We prove that, for a given UAV, \emph{have met (left)}
will be true after that UAV has experienced an event.  We then
establish that every UAV will experience an event within a time increment of $T$.
Consequently, after $T$ time increments, all UAVs will \emph{have
  met (left)}.  We then show that, if the i'th UAV's left neighbor is
left synchronized, the i'th UAV will be left synchronized in 1 time
increment or less.  Armed with this fact, and the fact the leftmost UAV is
always left synchronized, we prove by induction that all of the UAVs
in the ensemble will be \emph{left synchronized} within $(N)-1$ time
increments.  Combining the time to \emph{have met (left)} with the time to
\emph{left synchronized} gives us $2*T -1$ steps to left
synchronization.  The proof for \emph{have met (right)} and
\emph{right synchronized} follows from a symmetric argument.  Once we
know that a UAV satisfies both \emph{have met} and \emph{synchronized}
(both left and right), we can show that its behavior will be forever
periodic, establishing the convergence of the algorithm.

\subsection{Top-Level Theorem}

The top-level convergence theorem asserts that DPSS-A converges after
$2*T-1$ time increments.  Recall that the optimal surveillance pattern
requires that each UAV remain in its own segment.  This follows almost
immediately from the left and right synchronized predicates.  The
optimal pattern also requires that a UAV arrive at its segment
boundary at the same time as its neighbor (modulo perimeter
endpoints).  This follows from the synchronization predicates and the
fact that UAVs can only turn around when they meet other UAVs (or
reach a perimeter endpoint) and all UAVs travel at the same uniform
speed.  Our top-level theorem combines the concepts of bounded range
of motion and of synchronicity into a statement about the periodicity
of an arbitrary UAV.  It states that, following convergence, each UAV
will forever return to the same location in space every 2 time
increments (since time is normalized so that one time increment is the
amount of time it takes to traverse a segment once).  The following
predicate captures this notion of convergence, where {\tt one} is a
normalized time increment.




\begin{verbatim}
(defun-sk dpss-location-convergence (ens)
  (forall (i)
    (equal (UAV->location (ith-uav i (step-time (* 2 (one)) ens)))
           (UAV->location (ith-uav i ens)))))
\end{verbatim}

Our top-level theorem says that, after $2*T-1$ time increments, the
behavior of the ensemble satisfies our convergence predicate, where
the {\tt wf-ensemble} hypothesis captures our notion of a well-formed UAV
ensemble, (e.g., UAVs are sequentially ordered and on the perimeter).
We use {\tt (TEE)} in our formalization because {\tt T} is a reserved
symbol in the ACL2 package.

\begin{verbatim}
(defthm dpss-location-convergence-after-2T-1
  (implies
   (and
    (wf-ensemble ens)
    (step-time-always-terminates))
   (dpss-location-convergence (step-time (- (* 2 (TEE)) (ONE)) ens)))
  :hints (("Goal" :in-theory '(dpss-location-convergence-after-2T-1-helper))))
\end{verbatim}

The DPSS-A formalism consists of about 11K lines of ACL2 source code
and the complete proof of convergence requires about 30 minutes on a
standard laptop.

\section{ACL2 Proof Utilities}

The mechanization of the DPSS-A convergence proof was relatively
straightforward.  Every large proof effort, however, stresses existing
proof infrastructure.  Fortunately, ACL2 offers a number user
programmable techniques for tailoring and extending its automation to
address challenges in new domains.  Here we discuss three specific
ACL2 utilities, {\tt def::ung}, {\tt pattern::hint}, and {\tt def::linear},
that proved useful for specifying and reasoning about
the DPSS model.  While the {\tt def::ung} facility has been described
before in \cite{DEFUNG}, the {\tt pattern::hint} and {\tt def::linear} utility are
unique contributions of this work.

\subsection{{\tt \textbf{def::ung}}}

ACL2's definitional principle requires a proof of termination for
all recursive functions.  Note that the top-level DPSS-A simulator
{\tt step-time} is a recursive function that steps the DPSS system forward
by a specified quantity of time by taking some number of smaller
steps, each of which advances the system from one event to the next.
Technically, however, the time between events could be arbitrarily small.  In
other words: it could take an infinite number of events to consume any
given finite amount of time.  This observation highlights a challenge
in proving the termination of the step-time function.

One might argue that this scenario can't happen.  While some events
may happen arbitrarily close to one another, it would be impossible to
have more than a finite number of events in an arbitrarily small
quantity of time.  For example, if all the UAVs were clustered
together, very close, on one end of the perimeter, it could result in
a sequence of vanishingly small steps.  However, the ensemble will
eventually end up going in the other direction, toward the other end
of the perimeter, resulting a series much longer steps between events.
Since there are a finite number of UAVs all traveling at the same
uniform speed and only able to turn around at a perimeter endpoint,
when meeting a neighbor, or when ending an escort, we would expect
this to happen after no more than a finite number of small events,
likely bounded by the number of UAVs {\tt (N)}.

Based on the above argument, we believe that there is a proof that
{\tt step-time} always terminates.  However, the proof is likely to be ugly
and time consuming and it is unlikely that it will reveal anything of
interest about the DPSS algorithm or its convergence bound.
Consequently, it is unlikely that this proof would be of interest to
anyone studying DPSS.  In some sense, the need for this proof is
simply an artifact of the way we chose to model the behavior of the
system.

Fortunately, it is possible to delay termination proofs for recursive
functions by admitting them using the {\tt def::ung} macro \cite{DEFUNG}.  This macro
enables the admission of partial recursive functions (recursive
functions that may not terminate on all inputs) by extending the
proposed bodies of such functions with an additional domain check.
The domain check is a predicate which, if true, ensures that the
function terminates.  If the domain check fails, the modified function
returns a default value.  Otherwise, the body of the proposed
recursive function is evaluated.  Under the assumption that the
function always terminates (the domain check is always true), the
function admitted by {\tt def::ung} is exactly the same as the proposed
recursive function.  Better yet, the induction scheme it suggests is
also the same!  This means that any property that we might expect to
prove about the proposed recursive function can also be proved about
the function admitted by {\tt def::ung}.  Finally, we can formalize our
assumption that the function always terminates by simply defining a
universally quantified version of the domain check predicate generated
by {\tt def::ung}.



Currently the assumption that {\tt step-time-always-terminates} appears as an
explicit hypothesis in proofs about its behavior, including the final
proof of convergence.  Of course, given sufficient interest and
resources, a proper measure for {\tt step-time} could be developed and used
to dispatch this assumption, further strengthening our results.  In
the meantime, however, it has been useful to have a means of soundly
postponing the termination proof so that we could focus our efforts on
the aspects of the proof of most interest to the DPSS community.

\subsection{{\tt \textbf{pattern::hint}}}

The ACL2's proof philosophy emphasizes automation.  From induction to
rewriting, forward-chaining, and linear reasoning, ACL2 provides
automated support for proof steps that must be performed manually in
many other theorem proving systems.  Some reasoning steps, however,
are difficult to automate, even given ACL2's diverse set of
capabilities.  Fortunately, even when explicit user hints are
necessary, ACL2's computed hint facility provides a mechanism to
automate that as well.

Our DPSS model contains several lemmas and complex, quantified
formulae that can't be expressed efficiently as rewrite, linear, or
forward-chaining rules.  Some properties involve quantified relations
over more than one UAV.  Attempting to express the consequences of
such properties would, at best, involve expensive free variable
matching.  Making use of these lemmas, therefore, required us to
instantiate them by hand.  But the appropriate instantiations were
often unwieldy, difficult to predict, time-consuming to formulate,
and, of course, fragile and sensitive to change.

To address this issue, we developed the {\tt pattern::hint} facility.  This
facility extends ACL2's computed hints with a pattern matching
expression language that is evaluated against sub-goals.  For each
pattern match found, the resulting bindings are used to instantiate
hints that are passed to the theorem prover\cite{QUANT}.  The bindings are
then cached to avoid generating duplicate hints and a computed hint
replacement allows the process to continue to subsequent sub-goals.
Here is a simple example of a pattern match hint:

\begin{figure}[!htbp]
\begin{centerverbatim}
:hints ((pattern::hint  (<= x y) :use ((:instance helpful-lemma (a x) (b y)))))
\end{centerverbatim}
\end{figure}

This hint will search each sub-goal for instances of {\tt (<= x y)}
(yes, even though {\tt <=} is a macro) and, for each such instance,
will generate an instance of {\tt helpful-lemma} with {\tt a} bound to
{\tt x} and {\tt b} bound to {\tt y}.  Evaluating this pattern hint
against the following sub-goal:

\begin{figure}[!htbp]
\begin{centerverbatim}
(implies (and (< (foo x) 7) (<= x (foo x))) (< (foo a) (foo 7)))
\end{centerverbatim}
\end{figure}

Would result in a hint with two instances of helpful-lemma:

\begin{figure}[!htbp]
\begin{centerverbatim}
:use ((:instance helpful-lemma (a x) (b (foo x)))
      (:instance helpful-lemma (a (foo 7)) (b (foo a))))
\end{centerverbatim}
\end{figure}

The {\tt pattern::hint} framework helps to streamline and automate the
process of computing hints as a function of the goal.  In addition to
{\tt :use} hints, pattern hints also support lemma restrictions, case
splitting, function expansion, and theory manipulation.  The framework
features support for:

\begin{enumerate}
  \item Expressing patterns as untranslated terms, i.e., users don't need to worry about quoting constants or translating macros.
  \item Effectively matching against terms in the printed sub-goals (not the clause), i.e., the logical sense of the pattern matches that of the printed sub-goals, not the negated terms found in translated clauses.
  \item A variety of useful, composable pattern matching primitives.
  \item Instantiating hints with matched terms.
  \item Defining reusable, parameterized pattern functions and calling them by name.
  \item Defining and invoking reusable pattern hints by name.
\end{enumerate}

The following table provides a notional grammar for the supported pattern expressions:

\begin{center}
\begin{longtable}{| p{0.47\linewidth} | p{0.45\linewidth} |}
 \hline
 expr & Behavior \\
 \hline\hline
{\tt (:and . expr-list)} & Return the intersection of the bindings computed for expr-list, i.e., the returned bindings reflect a match for every element of expr-list. \\
 \hline
{\tt (:or . expr-list)} & Return the
union of the bindings computed for expr-list, i.e., the returned
bindings reflect a match for at least one element of
expr-list. \\
 \hline
{\tt (:first . expr-list)} & Return the first binding found for expr-list. \\
 \hline
{\tt (:match term expr)} & Instantiate term and pattern match it against expr. \\
 \hline
{\tt (:either term)} & Match either true or negated versions of term in the sub-goal. \\
 \hline
{\tt (:term term)} & Match term against any sub-term of the sub-goal. \\
 \hline
{\tt (:commutes expr symbol-alist)} & Match the expression and then treat the
pairs of symbols in symbol-alist as \emph{commuting variables} and
compute a binding that reflects every possible
commutation. \\
 \hline
{\tt (:replicate expr symbol-alist)} & Match the expression and then compute a Cartesian
product binding for each pair of symbols in symbol-alist. \\
 \hline
{\tt (:not expr)} & If expr does not match and produce a binding, continue. \\
 \hline
{\tt (:If expr expr expr)} & If the first pattern matches, continue with the
second, else evaluate the third. \\
\hline
{\tt (:implies expr expr)} & If the first
pattern matches, continue with the second else fail. \\
\hline
{\tt (:call (fn . term-list) symbol-list)} & Instantiate
each term in term-list from the current binding, call the specified
pattern function, and bind each symbol in symbol-list to the returned
values. \\
\hline
{\tt (:syntaxp term)} & Instantiate and evaluate the lisp term.  If not nil, continue. \\
\hline
{\tt (:equal . term-list)} & If all instantiated terms are syntactically equal, continue. \\
\hline
{\tt (:bind-free term symbol-list)} & Instantiate and evaluate the lisp term.  Bind each symbol in
symbol-list to the values returned.  Technically, term is expected to produce a list of bindings. \\
\hline
{\tt (:bind . term-alist)} & Bind each symbol to its associated instantiated term.
(May overwrite existing bindings.) \\
\hline
{\tt (:literally . symbol-list)} & Bind each symbol in the list to it's literal self. \\
\hline
{\tt (:keep . symbol-list)} & Filter the current binding, keep
only the symbols in symbol-list. \\
\hline
{\tt (:check term)} & Instantiate term and call the simplifier.  If term simplifies to true,
continue. \\
\hline
{\tt (:pass)} & Continue. \\
\hline
term & Instantiate term, treat it as a pattern, and match it against top-level terms in the sub-goal. \\
\hline
\end{longtable}
\end{center}

Some unique features of the pattern language include conditional
matching and meta-function support. The conditional operators
supported by the language include: {\tt (:not x)}, {\tt (:if a b c)},
and {\tt (:implies x y)}.  When we speak of \emph{conditions} what we
really mean is: did we find a match to the pattern?  The {\tt (:not x)} directive
says, essentially, continue if no match was found.  The {\tt (:if x
  y z)} directive says, if {\tt x} matches, then
proceed matching with {\tt y}, else proceed matching with {\tt z}.
The pattern language also supports the {\tt (:syntaxp )} and {\tt
  (:bind-free )} directives which enable the user to call-out to
arbitrary ACL2 functions to perform arbitrary computations on matched
terms.  The behavior of these operations is analogous to the behavior
of the related {\tt syntaxp} and {\tt bind-free} functions in ACL2
except that they don't support free variables like {\tt mfc} or {\tt
  state}.  If the {\tt :syntaxp} expression evaluates to {\tt nil}, it
acts as if no binding was found.  A function called within a {\tt
  :bind-free} expression is expected to return a list of bindings.

Performance was an issue with the DPSS proofs.  Most of the proofs
naturally involve many case splits.  Early attempts at automating
lemma instantiation often led to additional (unnecessary) case
splitting.  For example, we might have a lemma with a hypothesis like
{\tt (< x y)}.  However, the current clause might not know anything about
the relationship between the given instances of {\tt x} and {\tt y}.
As a result, ACL2 would case-split (often unnecessarily) into {\tt (<
  x y)} and {\tt (not (< x y))}.  To avoid this we wanted some
assurance, before instantiating such a lemma, that we could establish
{\tt (< x y)}.  We initially tried calling the theorem prover via {\tt
  (:check (< x y))} but we found this to be very expensive.  In the
end, we were able to use our pattern matching language to codify enough
meta-reasoning to establish such properties with high confidence.  For
example, if we could establish that {\tt x} is of the form {\tt (+ -3
  base)} and that {\tt y} is of the form {\tt (+ 2 base)}, then we
could safely conclude {\tt (< x y)}.

The framework also supports the definition of reusable, parameterized
pattern functions.  Such functions are defined using the {\tt
  def::pattern-function} macro.  Here we define a simple pattern
function, {\tt lte-match}, that takes no arguments and binds two
variables, {\tt a} and {\tt b}, such that, if successful, {\tt a} is
known to be less than or equal to {\tt b}.

\begin{figure}[!htbp]
\begin{centerverbatim}
(def::pattern-function lte-match () (:or (< a b) (<= a b)) :returns (a b))
\end{centerverbatim}
\end{figure}

This pattern function can then be used in other patterns via {\tt (:call)}:

\begin{figure}[!htbp]
\begin{centerverbatim}
(:call (lte-match) (x y))
\end{centerverbatim}
\end{figure}

It is also possible to define reusable named hints via {\tt def::pattern-hint}:

\begin{figure}[!htbp]
\begin{centerverbatim}
(def::pattern-hint lte-hint 
  (:call (lte-match) (x y)) 
  :use ((:instance linear-helper-lemma (a x) (b y))))
\end{centerverbatim}
\end{figure}

Such hints can then be referenced by name:

\begin{figure}[!htbp]
\begin{centerverbatim}
(defthm test (< (foo x) (goo y)) :hints ((pattern::hint lte-hint)))
\end{centerverbatim}
\end{figure}

Using named hints allowed us to define a handful of patterned hints
that could be used to properly instantiate several key lemmas.  Nearly
all the interesting DPSS proofs involve one or more pattern hints and
each pattern hint generates dozens to hundreds of instances of the key
lemmas over the course of each proof.  In total, pattern hints fired
986 times over the course of the entire proof and generated 1353 lemma
instances (not all pattern hints instantiate lemmas and some may
instantiate more than one lemma).

\subsection{{\tt \textbf{def::linear}}}

ACL2 incorporates specialized solvers for both linear and non-linear
arithmetic that enable it to decide many questions involving linear
relations over the standard arithmetic operations.  ACL2 also allows
the user to classify certain kinds of rules as {\tt :linear} rules,
extending ACL2's linear reasoning capability to include
user-defined functions.  ACL2 employs linear rules to automatically
add information to the so-called linear pot; the data structure
employed by the specialized linear solvers.  Tight integration between
the linear solvers and the rewriter is a crucial aspect of ACL2's
automated reasoning capabilities.

For example, a property such as:

\begin{figure}[!htbp]
\begin{centerverbatim}
(implies (< 0 x) (< (f x) x))
\end{centerverbatim}
\end{figure}

Could be designated as a linear rule and ACL2 would apply it
automatically to prove

\begin{figure}[!htbp]
\begin{centerverbatim}
(implies (< 0 x) (< (* 2 (f x)) (* 3 x)))
\end{centerverbatim}
\end{figure}

One of the properties of a well-formed DPSS ensemble is the fact that
the UAV locations are ordered.  In other words: a UAV with a lower ID
will never be located to the right of a UAV with a larger ID.
Formally, we would express this as:

\begin{figure}[!htbp]
\begin{centerverbatim}
(implies (< (UAV->id x) (UAV->id y)) (<= (UAV->location x) (UAV->location y))
\end{centerverbatim}
\end{figure}

Nearly every DPSS proof relies on this property at some point.  At
first blush, this property might appear to be an ideal candidate for a
{\tt :linear} rule.  The behavior we would want is, given two
instances, {\tt (UAV->location x)} and {\tt (UAV->location y)}, ACL2
should try to establish {\tt (< (UAV->id x) (UAV->id y))} and, if
successful, add {\tt (<= (UAV->location x) (UAV->location y))} to the
linear pot.  However, while ACL2 will accept this as a linear rule, it
generates an ominous warning about free variables.  If we dissect this
warning we learn that ACL2 intends to apply this rule by first
triggering on an instance of {\tt (UAV->location x)} and then
searching the type-alist for a term of the form {\tt (< (UAV->id x)
  (UAV->id y))} in hopes of finding a suitable binding for {\tt y}.
While this may work in some situations, it really isn't how we wanted
this rule to be applied!

Early in our development of the DPSS proofs, we solved this problem by
using {\tt pattern::hint} to simply instantiate our ordering lemmas as
needed.  This worked well enough, but was intellectually unsatisfying.
Our ordering properties look like linear rules, we should be able to
use them as such!  To finally assuage our frustration, we developed a
macro that massages a proposed linear rule of the general form
{\tt (implies (< x y) (< (f x) (f y)))} into a form that works as a linear
rule in the manner a user would most likely expect.  The key insight
was an observation by Matt Kaufmann that {\tt bind-free} could be used in
linear rules (just like with rewrite rules) to compute bindings for
free variables.  Consequently, a {\tt def::linear} event of the form:

\begin{figure}[!htbp]
\begin{centerverbatim}
(def::linear f-linear (implies (< x y) (< (f x) (f y)))
\end{centerverbatim}
\end{figure}

produces the following corollary:

\begin{figure}[!htbp]
\begin{centerverbatim}
(implies
  (and (bind-free (linear-binder ((f y)) (y) mfc state) (y))
       (< x y))
    (< (f x) (f y)))
  :trigger-terms ((f x))
\end{centerverbatim}
\end{figure}

\pagebreak

ACL2 will trigger this rule on an instance of {\tt (f x)}.  In the bind-free
hypothesis, the function {\tt linear-binder} will then search the linear pot
for terms that match the pattern {\tt (f y)} and return a list of candidate
bindings.  ACL2 then considers each binding in turn and, if it can
establish {\tt (< x y)}, it adds {\tt (< (f x) (f y))} to the linear pot.

This {\tt def::linear} macro was developed late in the process of
formalizing the DPSS proofs.  We had already been using {\tt
  pattern::hint} to find relevant instantiations of the ordering
property based on terms appearing in the clause.  While the {\tt
  def::linear} rules provide a more automated solution, this
automation comes at a cost.  In our experiments, the {\tt def::linear}
rules were noticeably more expensive than the {\tt pattern::hint}
solution.  This is likely because the {\tt pattern::hint} rules fire
late in the proof process, when the clause is stable and, crucially,
only when needed.  The {\tt def::linear} rules, however, fire
throughout the course of the proof, even when unneeded.  The final
proof actually uses a mix of {\tt pattern::hint} and {\tt
  def::linear}, favoring the former when performance is an issue and
the latter for simplicity.  The {\tt def::linear} and {\tt
  pattern::hint} macros can be found in the standard ACL2 books in the
files {\tt coi/util/linear.lisp} and {\tt
  coi/util/pattern-hint/pattern-hint.lisp} respectively.

\section{Conclusions and Future Work}

This report focused on efforts to mechanize in the logic of the ACL2
theorem prover a hand proof developed by Jeremy Avigad and Floris van
Doorn that bounds the convergence time of DPSS-A for an arbitrary
number of UAVs.  Our mechanized proof provides assurance that there
are no unfounded intuitions in the hand proof and provides confidence
that we have captured and addressed all essential assumptions and
corner cases related to the DPSS algorithm.  Furthermore, for each of
the two essential invariants employed in the hand proof, i.e. \emph{have met}
and \emph{synchronized}, we provide concise, useful
specifications relative to a concrete formalization of the DPSS-A
algorithm.  Finally, we discussed three ACL2 utilities that proved
useful in formalizing and reasoning about our DPSS model.

The DPSS proof scripts for the AvD proof are available as part of the
standard ACL2 books under {\tt projects/dpss/}.  Prior to learning of
the AvD proof, Collins was developing a proof of DPSS-A convergence
that employed several layers of abstraction and relied on the
characterization of certain emergent behaviors.  An incomplete version
of that proof has been released along with the AvD proof in the hope
that the abstract proof can be completed as well.  Our motivation for
pursuing a second, redundant proof is the hope that some of the
techniques it employs may eventually prove useful in analyzing the
more general DPSS Algorithm B (DPSS-B) which currently lacks a tight
convergence bound.

\pagebreak

\bibliographystyle{eptcs}
\bibliography{dpss}
\end{document}